\documentclass[apj]{emulateapj}  
\slugcomment{Accepted by the Astrophysical Journal Letters}

\newcommand{\snbf}{SN~2005bf}

\newcommand{\OxySeven}{\ion{O}{1}~$\lambda$7774}
\newcommand{\MgOne}{[\ion{Mg}{1}~$\lambda4571$}

\newcommand{\OxyOne}{[\ion{O}{1}]~$\lambda\lambda$6300,6363}

\newcommand{\OxyOneone}{[\ion{O}{1}]~$\lambda$6300}
\newcommand{\OxyOnetwo}{[\ion{O}{1}]~$\lambda$6363}
\newcommand{\CaTwo}{[\ion{Ca}{2}]~$\lambda\lambda$7291,7324}

\newcommand{\synCo}{$^{56}$\rm{Co}}

\newcommand{\kms}{\ensuremath{{\rm km~s}^{-1}}}

\newcommand{\snaj}{GRB~060218/SN~2006aj}

\newcommand{\snr}{SNR~1E0102.2$-$7219}

\newcommand{\nsn}{8}

\begin{document}
\title{Double-peaked Oxygen Lines Are not Rare in Nebular Spectra of Core-Collapse Supernovae}

\author{M.~Modjaz\altaffilmark{1,2,3}, 
R.~P.~Kirshner\altaffilmark{2},
S.~Blondin\altaffilmark{2},
P.~Challis\altaffilmark{2},
T.~Matheson\altaffilmark{4}
}

\altaffiltext{1}{University of California, Berkeley, 601 Campbell Hall, Berkeley, CA 94720; mmodjaz@astro.berkeley.edu}
 \altaffiltext{2}{Harvard-Smithsonian Center for Astrophysics,
 60 Garden Street, Cambridge, MA, 02138; kirshner,sblondin,pchallis@cfa.harvard.edu.}
\altaffiltext{3}{Miller Research Fellow}
\altaffiltext{4}{National Optical Astronomy Observatory, 950 N. Cherry Avenue, Tucson, AZ 85719-4933; matheson@noao.edu.}

\begin{abstract}

  Double-peaked oxygen lines in the nebular spectra of two peculiar
  Type Ib/c Supernovae (SN Ib/c) have been interpreted as off-axis
  views of a GRB-jet or unipolar blob ejections. Here we present
  late-time spectra of \nsn\ SN IIb, Ib and Ic and show that this
  phenomenon is common and should not be so firmly linked to
  extraordinary explosion physics. The line profiles are most likely
  caused by ejecta expanding with a torus- or disk-like geometry.
  Double-peaked oxygen profiles are not necessarily the indicator of a
  mis-directed GRB jet.

\end{abstract}

\keywords{gamma-ray burst: general supernovae: general}

\section{INTRODUCTION}\label{intro_sec}

To understand the connection between long-duration Gamma Ray Bursts
(GRBs) and the supernovae (SNe) associated with them, it would be
significant to detect the effects of jets that are produced in the
explosion even when the observer is not in the relativistic
beam \citep{woosley06_rev}). In the broad-lined Type Ic, SN 2003jd, an
off-axis GRB jet was proposed as the cause for the double-peaked
oxygen profile observed in its nebular spectrum
\citep{mazzali05_03jd,valenti08}. A similar double-peaked oxygen
profile, with a large blueshift, was detected in the peculiar Type Ib,
SN 2005bf \citep{maeda07,modjaz07_thesis}, where it was interpreted as
coming from a unipolar blob or jet \citep{maeda07}.

If the double-peaked oxygen shape were exclusively caused by highly
relativistic jets, then we might expect only broad-lined SN Ic, whose line
widths approach 30,000 \kms , and which are the only subtype of
stripped-envelope SN seen at the sites of GRBs to exhibit
them. However, if we see double-peaked profiles in a range of
supernova types, we might conclude that asphericities are present in
more typical core-collapse events.

We set out to obtain a set of nebular spectra for supernovae of
various subtypes to measure the emission line profiles with adequate
resolution and signal to detect signs of these double-peaked profiles
in nebular spectra of supernovae. As the supernova turns optically
thin a few months after maximum light, the emission line shapes can
provide information on the velocity distribution of the ejecta
\citep{matheson01,foley03,maeda06}. Structure in emission line
profiles have been reported in SN II \citep{spyromilio91}, SN IIb
\citep{spyromilio94_93j,matheson00_93jdetail} and SN Ib
\citep{sollerman98,elmhamdi04}, and interpreted as clumping of oxygen
in the ejecta.

In \S~\ref{obs_sec} we describe the observational sample. In
\S~\ref{doubleOxy_sec} we discuss the line profiles and show that the
double-peaked line profiles are unlikely to be caused by optical depth
effects. We speculate on the implications of our results in
\S~\ref{interpretation_sec} and summarize in \S~\ref{conclusion_sec}.

\section{Observations and Analysis}\label{obs_sec}

Spectra were obtained with the 6.5 m Clay Telescope of the Magellan
Observatory located at Las Campanas Observatory, the 8.1 m
Gemini-North telescope via queue-scheduled observations (GN-2005B-Q-11,
GN-2006B-Q-16 PI: Modjaz), the 6.5 m Multiple Mirror Telescope
(MMT) and the 1.5 m Tillinghast telescope at the Fred Lawrence Whipple
Observatory (FLWO). The spectrographs utilized were the LDSS-3
(Mulchaey \& Gladders 2005) at Magellan, the GMOS-North \citep{hook03}
at Gemini, the Blue Channel \citep{schmidt89} at the MMT, and FAST
\citep{fabricant98} at the FLWO 1.5 m telescope. All optical spectra
were reduced and calibrated employing standard techniques in
IRAF\footnote{IRAF is distributed by the National Optical Astronomy
  Observatory, which is operated by the Association of Universities
  for Research in Astronomy, Inc., under cooperative agreement with
  the National Science Foundation.} and our own IDL routines for flux
calibration (see e.g., \citealt{matheson08}).

\begin{figure}[!ht]                
\includegraphics[scale=0.55,angle=0]{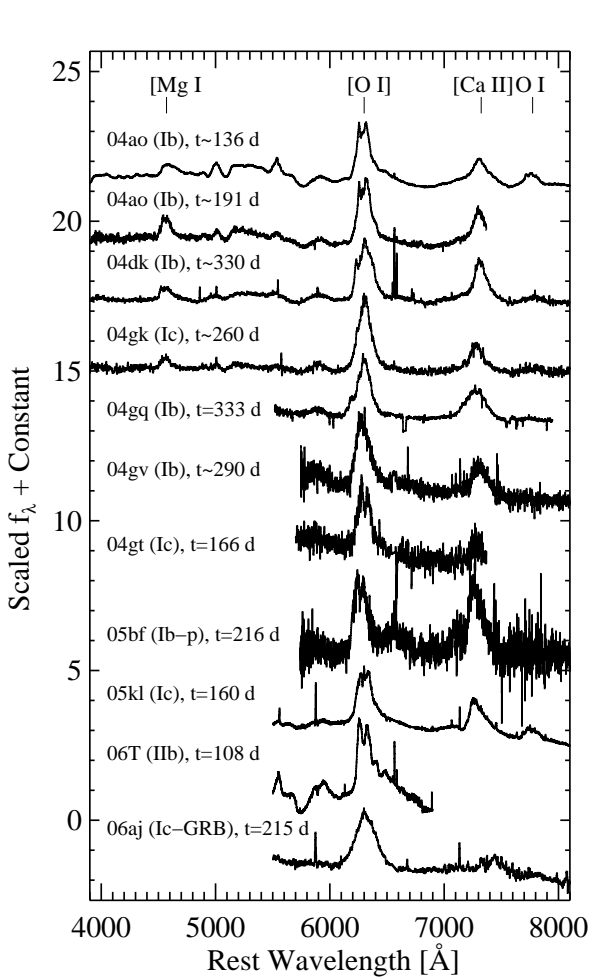}
\singlespace \caption{Selection of nebular spectra of SN IIb, SN Ib
  and SN Ic in their respective rest frames. SN name, type and phase
  of spectrum (with respect to maximum light, except for \snaj\ which
  is referenced to time of GRB burst) are marked. For clarity we show
  only the latest spectrum of each SN (except for SN~2004ao), even
  though we obtained several for some SN. Also the main nebular
  emission lines of \MgOne, \OxyOne , \CaTwo\ and \OxySeven\ are
  marked at the very top.  }
\label{nebmont_fig}
\end{figure}                                                                 

In Figure~\ref{nebmont_fig}, we present the observed nebular spectra
of \nsn\ new SN IIb, SN Ib and SN Ic. For clarity we show only the
latest spectrum of each SN (except for SN~2004ao). We double the
number of unambiguous SN Ib with late-time data from four historical
cases (SNe 1983N, 1984L, 1990I, and 1996N) to a total of eight (now
including SNe~2004ao, 2004dk, 2004gq, 2004gv), not counting SN~2005bf
(see also \citealt{maeda08}). We determined the phase with respect to
maximum from our own photometry (\citealt{modjaz07_thesis}, M.
Modjaz, in prep.). In addition, we show our data of the peculiar SN Ib
2005bf, independently observed and analyzed by \citet{maeda07}, and of
SN~2006aj, the broad-lined SN Ic connected with GRB~060218
\citep{modjaz06}, independently observed and analyzed by
\citet{mazzali07_06aj}.

All spectra display the hallmark features of nebular SN Ib/c spectra:
strong forbidden emission lines of intermediate-mass elements such as
\OxyOne\ and \CaTwo, similar to spectra of SN~II.  In the absence of
hydrogen, oxygen is the primary coolant in the ejecta of
stripped-envelope SN at late epochs when the gas is neutral or at most
singly ionized \citep{uomoto86,fransson87}. SN~2005bf is the only SN
discussed here that exhibits broad H$\alpha$ (see also
\citealt{maeda07}).

\section{Detection of Double-peaked Oxygen Lines }\label{doubleOxy_sec}

In the optically-thin case, the late-time emission line profile is
dictated, in principle, by the geometry and distribution of the emitting
material \citep{fransson87,schlegel89}. A radially expanding spherical
shell of gas produces a square-topped profile, while a filled uniform
sphere produces a parabolic profile. In contrast, a cylindrical ring,
or torus, that expands in the equatorial plane gives rise to a
``double-horned'' profile as there is very little low-velocity
emission in the system, while the bulk of the emitting gas is located
at $\pm v_t$, where $v_t$ is the projected expansion velocity at the
torus.

\begin{figure}         
\includegraphics[scale=0.55,angle=0]{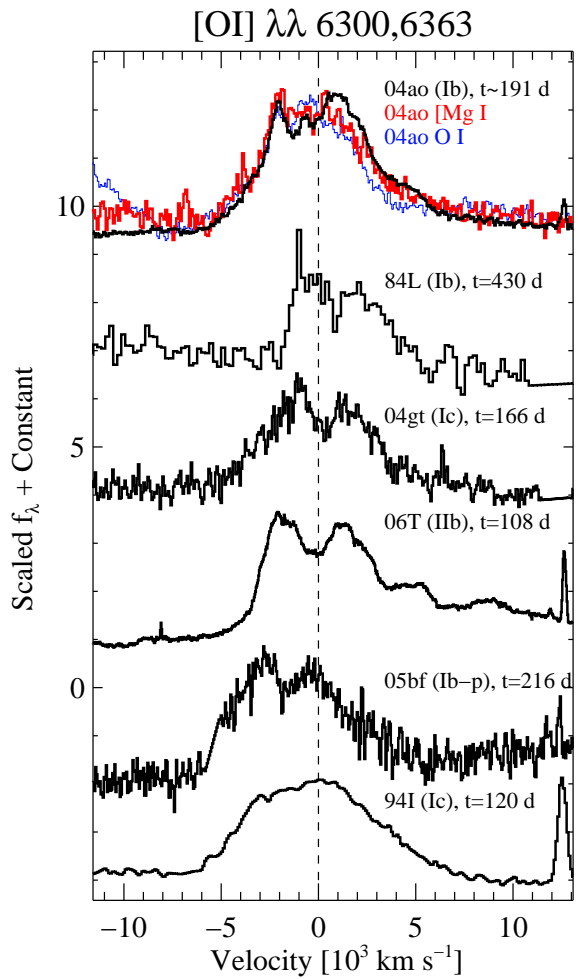}
\singlespace \caption{Montage of SN with double-peaked oxygen profiles
  in velocity space. SN name, type and phase of spectrum (with respect
  to maximum light).  SN~1984L, a SN Ib is from \citet{schlegel89}.
  SN~1994I \citep{filippenko95} that exhibits a simple parabolic
  oxygen line profile is plotted for comparison at the bottom. The
  dashed line marks zero velocity with respect to 6300 \AA\ . For
  SN~2004ao, we plot the scaled profiles of \OxySeven\ (in blue) and
  \MgOne\ (in red), which are not doublets, but also exhibit the two
  peaks. As discussed in the text (\S~\ref{optdepth_subsec}), the two
  horns are unlikely to be due to the doublet nature of \OxyOne.  Note
  the large blueshift in \snbf\ compared to the other SN which we do
  not explain here.}
\label{doubleOxy_fig}
\end{figure}                                                                 
Figure~\ref{nebmont_fig} shows that SNe 2004ao, 2004gt, 2006T show
conspicuous double-peaked lines of \OxyOne . We plot the oxygen line
profiles in velocity space for these objects in
Figure~\ref{doubleOxy_fig}. A literature search of published SN IIb,
Ib, and Ic spectra reveals that SN Ib 1984L, the prototype of the SN
Ib class, also shows a clear double-peaked profile that went
unremarked in the original publication \citep{schlegel89}. We plot
the spectrum with the highest S/N in Figure~\ref{doubleOxy_fig}. Finally,
we also include our data on SN~2005bf, the peculiar SN Ib with
extraordinary early-time light curves and spectra
\citep{tominaga05,folatelli06}, that showed highly blue-shifted (by
$\sim$ 2000 \kms) oxygen and calcium lines
\citep{maeda07,modjaz07_thesis}.

This double-peaked line profile is visible in the oxygen line and not
in calcium. For SN~2004ao, the only SN for which we have a
sufficiently high S/N spectra, the double-horned feature is present in
the permitted line of oxygen, \OxySeven , and in \MgOne\ (see
Fig.~\ref{doubleOxy_fig}). Our late-time spectral follow-up for some
SN \citep{modjaz07_thesis} shows that SN \emph{without} double-peaked
profiles (SNe 2004gk, 2004gq) do not exhibit them at any epoch, while
SN \emph{with} double peaks (SNe~2004ao, 2004gt) retain them over the
full observed period.

The two horns are roughly symmetrically offset by $\sim$1000$-$2000
\kms\ around the trough. For most SN in our sample, the trough between
the two ``horns'' is located at nearly zero velocity for [\ion{O}{1}]
6300\AA. SN~2004ao shows an apparent blueshift and SN~1984L shows a
redshift in the trough. A velocity shift of $\pm$ 200 \kms\ (as seen for
SN 2004gt) could be due to rotation of the host galaxy. We discuss other
possible causes for these shifts in \S~\ref{interpretation_sec}.

\subsection{Optical Depth Effects causing the Double Peaks? }\label{optdepth_subsec}

\begin{figure}[!ht]         
\includegraphics[scale=0.37,angle=90]{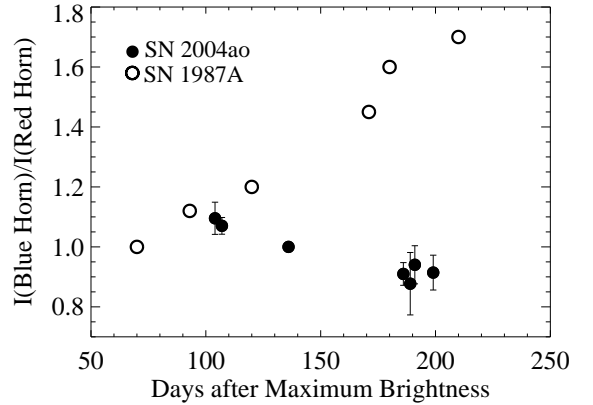}
\singlespace \caption{Evolution of the intensity ratio of the blue to
  the red horn ( I(Blue Horn)/I(Red Horn) ) as a function of time for
  SN~2004ao (filled circles, this work). The intensity ratio decreases
  over time.  In contrast, if the blue and red horns corresponded to
  \OxyOneone\ and \OxyOnetwo\, the intensity ratio would increase over
  time, as observed in SN~1987A (empty circles, from \citealt{li92})
  and 1988A \citep{spyromilio91}.  }
\label{sn04aoratio_fig}
\end{figure}

Could the observed double-peaked profile be due to optical depth
effects?  The \OxyOne\ doublet has a velocity separation of 3000 \kms
(with respect to 6300 \AA) and an intensity ratio of 3:1 in the
optically thin limit. Optical depth effects for \OxyOne\ have been
discussed in the supernova context by \citet{leibundgut91},
\citet{spyromilio91} and \citet{li92}. In these models, the ratio of
\OxyOneone\ to \OxyOnetwo\ evolves from $\sim$1 to $\sim$3 from early
times to late times as the supernova expands and the lines become
optically thin, as was observed for SNe 1987A
\citep{spyromilio91_87a,li92} and 1988A \citep{spyromilio91}. In
Figure~\ref{sn04aoratio_fig}, we plot the measured ratio of the blue
to the red horn for SN 2004ao, the supernova in our sample with the
longest time span of observations. The ratio of line intensity
decreases over time for SN~2004ao. In contrast, this ratio would
increase with time if the two horns were due to \OxyOneone\ and
\OxyOnetwo , as seen in SN 1987A (\citealt{li92}, overplotted in
Fig.~\ref{sn04aoratio_fig}).

In SN 2004ao, where we have sufficiently high S/N- data, we also see a
similar double-peaked profile in the permitted \OxySeven\ line and in
\MgOne\, which are not doublets, as shown in Fig.~\ref{doubleOxy_fig}.
This suggests that the line shapes really do provide a guide to the
distribution of oxygen and magnesium in the ejecta. Optical depth is
not the probable cause for seeing two equal peaks in the \OxyOne\ line
profile.

\section{Interpretation}\label{interpretation_sec}

We suggest a torus-like structure for either the oxygen distribution
or for the radioactive energy source that excites the oxygen lines,
\synCo\ \citep{fransson89}. If the distribution of the energy source
were most important, we would expect to see a double-peaked profile
for all the emission lines, including calcium.  Since the
double-peaked lines are seen only for oxygen and magnesium, the
evidence favors the view that we are seeing the effects of the
distribution of those elements in velocity space.

Models of the line shape that results from an expanding disk are
worked out by, e.g., \citet{gerardy00}, and \citet{fransson05}.
Comprehensive modeling of the line profiles needs to take into account
viewing angle effects and the thickness of the line-emitting and
expanding region.  We leave that as a topic for future work.

Here we do not attempt to explain the extreme blueshift ($\sim$ 2000
\kms) seen in \snbf\ at epochs later than 200 days. The less extreme
blueshifts seen in spectra obtained at $t \la$ 200 days for one of our
SN (SN~2004ao) might be caused by the same mechanism that is
responsible for the observed blueshifts in SN that show a single-peak
line of \OxyOne . S. Taubenberger at al. (in preparation) observe in a
large set of late-time spectra of 34 SN Ib and SN Ic that the oxygen
line centroids are found to be blueshifted for spectra taken at $t
\la$ 200 days. These blueshifts range up to $\sim$1500 \kms\ for
spectra at $t\sim$90 days and go to zero with increasing time.
Taubenberger et al. exclude dust formation, contamination from oth er
lines and geometric effects as potential causes and invoke residual
opacity effects as the most likely reason.  The observed
blueshift-values in our sample agree with those seen in Taubenberger
et al. for similar epochs; thus, the same mechanism might be causing
the blueshifts in both single- and double-peaked \OxyOne . We
encourage future detailed modeling of the radiative transfer of
\OxyOne\ to elucidate the exact reason for the blueshift. We do not
have a simple explanation for the redshift seen in the trough of the
double-peaked \OxyOne\ of SN~1984L.

We conclude that the most probable explanation for the double-horned
oxygen profiles is a torus like, or at least flattened, distribution
of oxygen, that may be a relic of the explosion physics. Energy
deposition and radiation transport modeling of late-time emission in
SN Ib (\citealt{fransson89}) predicts that material contributing to
\OxyOne\ (and to \MgOne) emission is situated further out in the
ejecta than material emitting in \CaTwo\ (see their Fig.  8). We
speculate that the mechanism causing the asphericities has to affect
the outer layers of the SN debris more than the inner ones.  The
anisotropies we observe in the oxygen lines suggest large-scale plumes
of mixing rather than the small-scale inhomogeneities invoked to
explain sub-structure of the oxygen lines as reported for SNe II
1979C, 1980K (\citealt{fesen99} and references therein), in SN IIb
1993J \citep{matheson00_93jdetail} and in SN Ib/c~1985F
\citep{filippenko86}. In the case of our observed SN, the anisotropies
may be of global nature in order to give rise to such a clear
double-peaked line profiles. Alternatively, in case the supernova
progenitors are part of binaries, binary interaction or merger might
be modulating the geometry of the supernova ejecta \citep{morris07}.

Such a clear signature of a pure double-peaked oxygen profile has not
been found in other SN Ib/c \citep{matheson01}, except in SN~2003jd
\citep{mazzali05_03jd} and in SN~2005bf. SN~2003jd was a broad-lined
SN Ic without an observed GRB and showed the same double-peaked oxygen
profile \citep{mazzali05_03jd} as SN~2004ao (compare their Fig. 2 with
our Fig.~\ref{doubleOxy_fig}). \citet{mazzali05_03jd} interpret their
observations as indicating an aspherical axisymmetric explosion viewed
from near the equatorial plane, and suggest that this asphericity was
caused by an off-axis GRB jet. They conclude that only special SN,
those with high expansion velocities and possibly connected with GRBs,
are aspherical, and use SN~2003jd as a link between the normal,
spherical SN Ic and those highly aspherical ones connected with GRBs.
In our sample, however, the SN showing these line profiles are normal
SN Ib and even SN IIb, i.e. SN from stars with intact helium and
partial H envelopes before explosion, and have normal early-time
expansion velocities. Thus, it appears that asphericities are
prevalent in normal core-collapse events.

Indeed, complex and ring-like velocity structures of oxygen have been
observed in SN remnants (SNRs), most prominently in the oxygen-rich
\snr\ in the Small Magellanic Cloud \citep{tuohy83}. \citet{tuohy83}
and subsequent papers fit a twisted ring model to the spatial and
velocity extent of the filaments that extents to velocities ranging
between $-$2500 to $+$ 4000 \kms. It is conceivable that the stellar
death leading to \snr\ had a geometry for the ejecta that was similar
to the events we observe.  Asphericities may be common in in
core-collapse events, be they neutrino-driven \citep{scheck06},
acoustic \citep{burrows06}, or magneto driven
\citep{burrows07,dessart08}, as indicated by polarization, neutron
star kick velocities and the morphologies of young remnants.
Recently, \citet{maeda08} also reported that double-peaked line
profiles are not rare in stripped core-collapse SN. They suggested
that the observed fraction of double-peaked profiles ($\sim40\pm10$\%)
is consistent with the hypothesis that all of these core-collapse
events are mildly aspherical. Although their inferences are based on
the results of bipolar jet models \citep{maeda06}, \citet{maeda08}
suggest that a wide variety of geometries, similar to those suggested
here, may be present.

\section{Conclusions}\label{conclusion_sec}

In summary, we detect clear double-peaked line profiles of \OxyOne\ in
three SN IIb and SN Ib/c out of our observed sample of \nsn\ and
additionally, in one out of four published SN Ib. Our sample does not
include broad-lined SN Ic or peculiar objects.  These line profiles
are probably not caused by optical depth effects, and suggest global
anisotropies in the ejecta. Prior to this work, double-peaked oxygen
lines had only been reported in two peculiar SN Ib/c and interpreted
as off-axis GRB-jet or unipolar blob ejections. It seems more likely
that asphericities are present in a wide variety of core-collapse
events and they are not strictly confined to the supernovae associated
with GRB jets. Although special models have been proposed to account
for the line profiles in peculiar supernovae, our investigation
suggests that double-peaked profiles, and underlying disks, are not
unusual. These ordinary SN from our sample have a variety of subtypes,
reflecting the diversity of mass loss prior to the explosion.

We recommend a meta-analysis of available nebular spectra of all
core-collapse SN, spanning the full range from SN II to IIb, Ib, Ic
and finally, to broad-lined SN Ic, in order to quantify the kinematics
and geometry of the ejecta and to seek trends as a function of SN
type. Further high-SN/N multi-epoch observations of a sample of SN
II/IIb/Ib/Ic coupled with radiative transfer models should help to
elucidate the observed blue- and redshifts of the line profiles.


\acknowledgements M. M. would like to especially thank R. Fesen for
very helpful suggestions and for reading a draft of this manuscript,
and in addition C. Fransson, R. Narayan, D. Sasselov, B. Leibundgut
and J.  Spyromilio for insightful discussions. We thank and FLWO 1.5m
observers for obtaining service-spectroscopy, and the referee, Alex
Conley, for constructive comments. M. M. acknowledges support from the
Miller Institute for Basic Research during the time in which part of
this work was completed. Observations reported here were obtained at
the MMT Observatory, a joint facility of the Smithsonian Institution
and the University of Arizona, at the 6.5 meter Magellan Telescopes
located at Las Campanas Observatory, Chile, at the F.L Whipple
Observatory, which is operated by the Smithsonian Astrophysical
Observatory, and at the Gemini Observatory, which is operated by the
Association of Universities for Research in Astronomy, Inc., under a
cooperative agreement with the NSF on behalf of the Gemini
partnership. Supernova research at Harvard University has been
supported in part by the National Science Foundation grant AST06-06772
and R.P.K. in part by the Kavli Institute for Theoretical Physics NSF
grant PHY99-07949.

{\it Facilities:}  \facility{MMT (Blue Channel spectrograph)}, \facility{FLWO:1.5m (FAST)}, \facility{Magellan:Baade(LDSS3)}, \facility{Gemini:Gillett (GMOS-N)}

\end{document}